# Synthesis, Structural, and Transport Properties of Cr-doped BaTi$_2$As$_2$O

Qiucheng Ji,[a] Yonghui Ma,[a] Kangkang Hu,[a,b] Bo Gao,[a] Gang Mu,[a,]* Wei Li,[a,c] Tao Hu,[a] Ganghua Zhang,[d] Qingbiao Zhao,[d] Hui Zhang,[d] Fuqiang Huang,[d] and Xiaoming Xie[a]

[a] *State key Laboratory of Functional Materials for Informatics and Shanghai Center for Superconductivity, Shanghai Institute of Microsystem and Information Technology, Chinese Academy of Sciences, Shanghai 200050, China*

[b] *College of Sciences, Shanghai University, Shanghai 200444, China*

[c] *State Key Laboratory of Surface physics and Department of Physics, Fudan University, Shanghai 200433, China*

[d] *CAS Key Laboratory of Materials for Energy Conversion, Shanghai Institute of Ceramics, Chinese Academy of Sciences, Shanghai 200050, China*

* Corresponding author. Fax: +86-21-62127493. Email: mugang@mail.sim.ac.cn

**Abstract**: The interplay between unconventional superconductivity and the ordering of charge/spin density wave is one of the most vital issues in both condensed matter physics and material science. The Ti-based compound BaTi$_2$As$_2$O, which can be seen as the parent phase of superconducting BaTi$_2$Sb$_2$O, has a layered structure with a space group P4/mmm, similar to that of cuprate and iron-based superconductors. This material exhibits a charge density wave (CDW) ordering transition revealed by an anomaly at around 200 K in transport measurements. Here, we report the synthesis and systematical study of the physical properties in Cr-doped BaTi$_{2-x}$Cr$_x$As$_2$O (x = 0 - 0.154), and demonstrate that the transition temperature of the CDW ordering is suppressed gradually by the doped Cr element. The magnetization measurements confirm the evolution of the CDW ordering transition. These observed behaviors are similar to that observed in iron-based superconductors, but no superconductivity emerges down to 2 K. In addition, the first-principles calculations are also carried out for well-understanding the nature of experimental observations.

Keywords: Cr-doped BaTi$_2$As$_2$O, Charge-density-wave (CDW), Superconductivity

## 1. Introduction

Since the discovery of cuprate and iron-based superconductors,[1,2] the search for high-temperature superconductivity and novel superconducting mechanisms has become one of the most challenging tasks of condensed matter physicists and material scientists. From the viewpoint of theoretical and experimental evidences,[3-5] these superconductivities often emerge in the proximity of the ordering of spin/charge density wave instabilities, in contrast to that of conventional BCS superconductivity[6]. In recent years the studies of BaTi$_2$Pn$_2$O (Pn = As, Sb, Bi) system associated with charge density wave instabilities have attracted great research attentions since the discovery of superconductor $Ba_{1-x}Na_xTi_2Sb_2O$ ($T_C^{onset}$ =5.5 K).[7] The crystal structure of $BaTi_2Pn_2O$ combines the features of cuprate and iron-based superconductors (see Figure 1(a)): It consists of edge-shared [$Ti_2Pn_2O$] layers separated by the layers of $Ba^{2+}$. The conductive layer is the [$Ti_2Pn_2O$] layer, where the atoms show a unique connective feature. It has a $Ti_2O$ square net with an anti-$CuO_2$-type structure of the cuprate superconductors, where the Pn atoms are located above and below the $Ti_2O$ squares. Four Ti atoms and two Pn atoms form the octahedral sharing the corners with the neighboring one.

Interestingly, an anomalous phase transition has been observed in both the Sb- and As-based compounds by the susceptibility and resistivity measurement.[8-12] Such an anomaly was ascribe to the charge density wave order.[13] The substitution of $Ba^{2+}$ by $Na^+$ in $BaTi_2Sb_2O$ as hole doping leads to the suppression of charge density wave ordering phase and the emergence of superconductivity with highest superconducting transition temperature $T_c$=5.5 K when x ≈ 0.15.[7, 14, 15] On the other hand, X. H. Chen *et.al.* successfully synthesized $BaTi_2As_2O$ in 2010, and increased the transition temperature of charge density wave order to 200 K.[16] It was reported that element doping is very difficult and only $Li^+$ has been doped into interstitial sites of $BaTi_2As_2O$. But no superconductivity was observed. Therefore, more efforts are needed in this system to explore the possible superconductivity and to study the nature of the interplay between superconductivity and CDW order.

In this work, chromium element was doped into BaTi$_2$As$_2$O and their physical properties were fully investigated. We show the evolution of CDW ordering transition in BaTi$_{2-x}$Cr$_x$As$_2$O as a function of the substitution concentration. Evidences are displayed to demonstrate that, with the increase of chromium content, the CDW ordering transition temperature (T$_{CDW}$) is continuously lowered, but without the emergence of superconductivity down to 2 K. In addition, we carried out the first-principles calculations for both parent and doped compounds for well-understanding the nature of our experimental observations.

## 2. Results and discussions

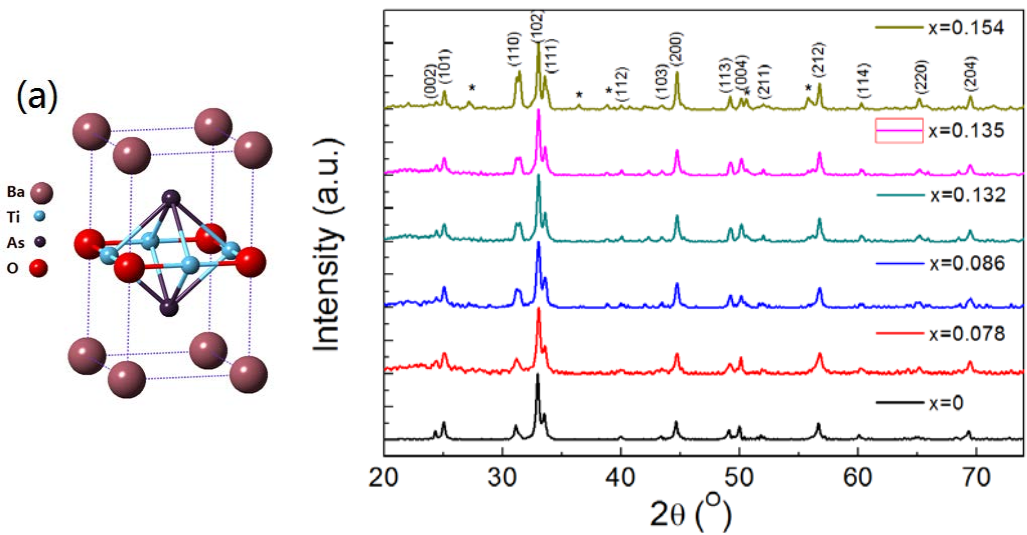

**Figure 1.** (a) Structure of BaTi$_2$As$_2$O. (b) X-ray diffraction patterns of BaTi$_{2-x}$Cr$_x$As$_2$O with different doping levels. Main peaks can be indexed

to the layered structure of $BaTi_{2-x}Cr_xAs_2O$. Small amounts of impurities are marked by the asterisks.

The actual doping levels of the samples were analyzed and determined by the energy dispersive X-ray spectroscopy (EDS) measurements. The results can be seen in the supporting information and Figure S1. The values (x) determined from EDS measurements will be used in the present paper. We also examined the structure and purity of our samples by the powder X-ray diffraction (XRD) measurements at room temperature. In Figure 1(b) we show the XRD data of the $BaTi_{2-x}Cr_xAs_2O$ samples with $0 \leq x \leq 0.154$. Main diffraction peaks can be indexed to the layered structure with the space group P4/mmm as shown in Figure 1(a). Tiny peaks from small amounts of impurities can be detected as the doping level increases. Moreover, we can see that the peaks move to the right slightly with the increase of doping, indicating the doping induced shrinkage of crystal lattice. To investigate the influences on the crystal lattice by the Cr substitution quantitatively, we obtained the lattice parameters by fitting the XRD data. The a- and c-axis lattice parameters and volume decrease gradually. The detailed descriptions and discussions about the response of crystal lattice to Cr doping are presented in the supporting information and Figure S2. The clear evolution of the crystal lattice suggests the successful substitution of Cr element.

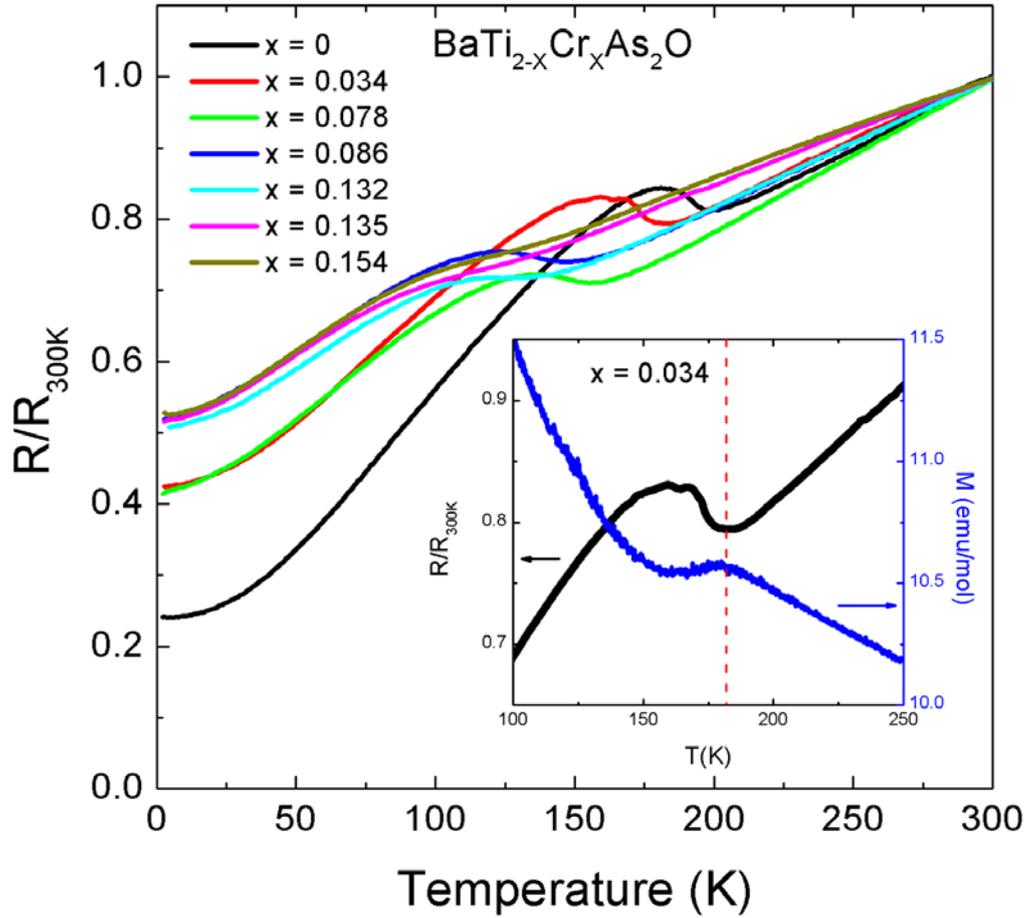

**Figure 2.** Temperature dependence of the resistance for $BaTi_{2-x}Cr_xAs_2O$ samples, normalized to the data at 300K. Inset is the comparison of R-T and M-T curves for one typical sample with x = 0.034 to confirm the CDW transition at $T_{CDW}$.

Resistance and magnetism measurements were performed to study the transport and possible superconducting properties. Figure 2 indicates the temperature dependence of normalized resistance of our samples with doping $0 \leq x \leq 0.154$. A clear anomaly is observed around 200 K for the sample with x = 0, which is consistent with the previous report.[16] This anomaly has been identified as a nematic CDW ordering transition.[13] For the samples in low doping region, such a transition was also detected by the magnetism measurements. Taking the sample with x= 0.034 as an example, the transition temperature $T_{CDW}$ determined in the M-T curve is rather consistent with that in the R-T curve, as shown in the inset of Figure 2. As more and more chromium is doped into the sample, the CDW transition temperature declines gradually. But no superconductivity emerges down to 2 K.

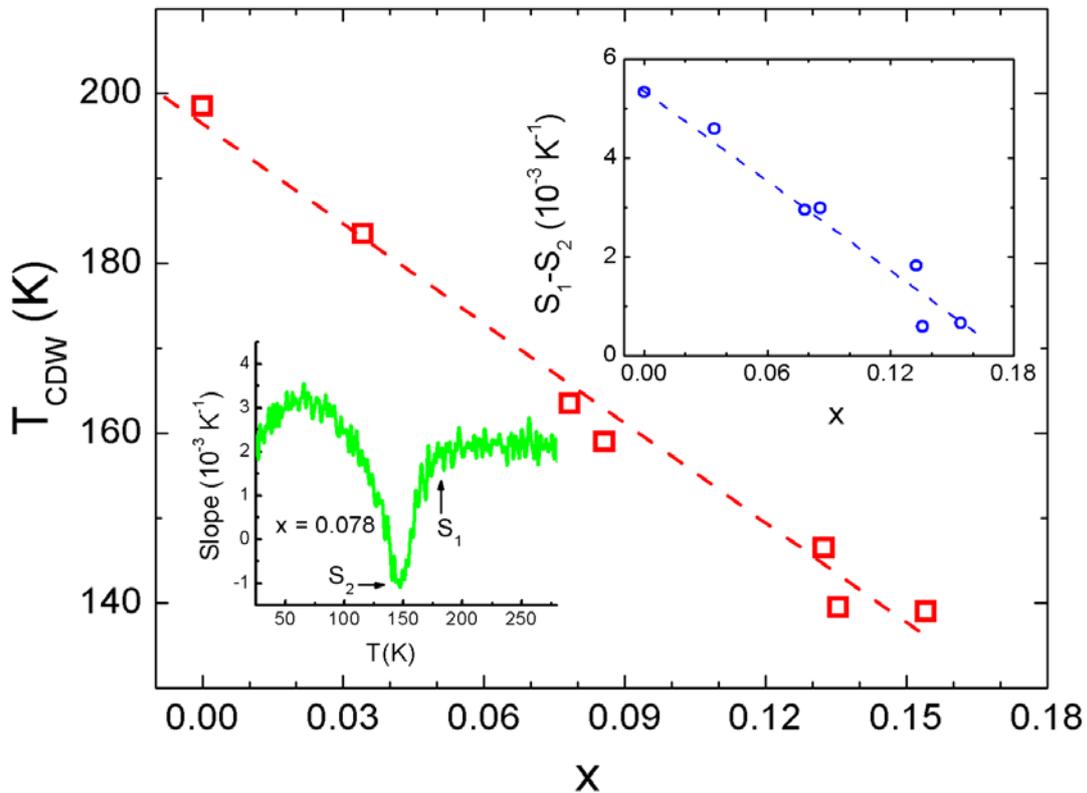

**Figure 3.** Doping dependence of CDW transition temperature $T_{CDW}$. The dashed line is the guide for eyes. The slope curve of R-T curve for the sample with x=0.078 is plotted in the left inset. Doping dependence of

$S_1$-$S_2$ is shown in right inset.

In Figure 3, we show the doping dependence of the CDW transition temperature $T_{CDW}$. The dashed line is a guide for eyes. $T_{CDW}$ goes down gradually and shows a roughly linear tendency with the increase of doping concentration. The lowest $T_{CDW}$ obtained from our experiment is 139 K. Further suppression of $T_{CDW}$ is frustrated due to the solubility limit of Cr and the indistinctive resistive features from the CDW transition in the high doing region. Here, we also attempted to describe such resistive features of the CDW transition quantitatively. As shown in the left inset of Figure 3, the temperature derivative of resistance, $Slope = \frac{d(R/R_{300K})}{dT}$, for the sample with x = 0.078 is plotted versus temperature. In this figure, the onset point of the transition $S_1$ and the minimum point $S_2$ are determined. The difference of slopes between the two points $S_1$-$S_2$ is used to describe the feature due to the CDW transition. In the right inset of Figure 3, we show the doping dependence of $S_1$-$S_2$. The value of $S_1$-$S_2$ also decreases linearly with the doping level x and becomes nearly zero with x around 0.15, which suggests that the features due to the CDW ordering are almost unobservable in the resistance data in the high doping region.

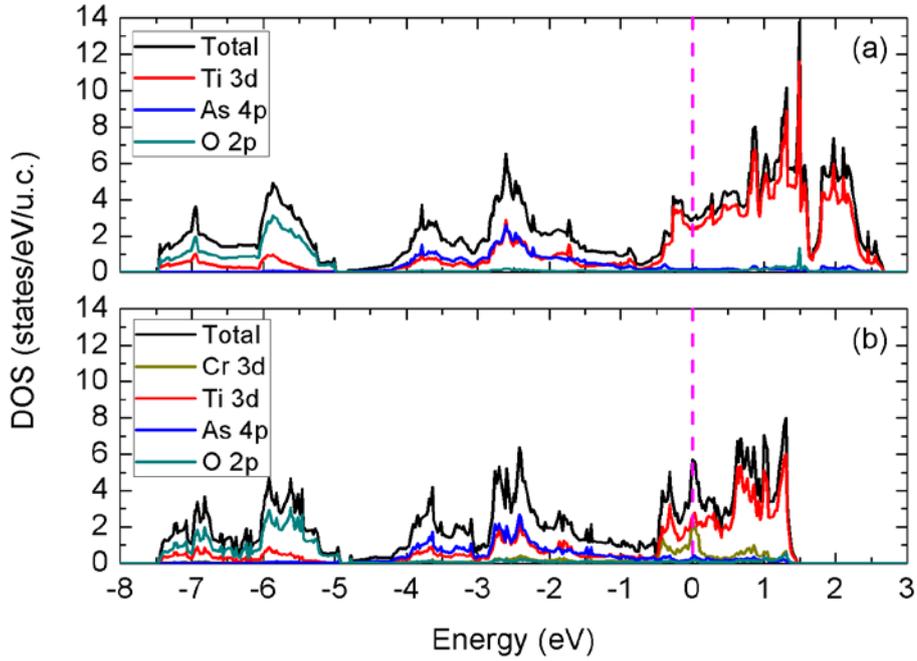

**Figure 4.** The total and projected DOS on Ti 3d (Cr 3d), As 4p, and O 2p orbitals per unit cell of (a) $BaTi_2As_2O$ and (b) $BaTi_{1.75}Cr_{0.25}As_2O$ in the nonmagnetic state by using a GGA calculations. The Fermi energy is set to zero (dish line).

In order to well-understand the nature of experimental observations, we carry out the first-principles calculations[17-20] to discuss the evolution of electronic structure as Cr doped in $BaTi_2As_2O$. The calculated total and projected density of states (DOS) for $BaTi_2As_2O$ and Cr doped $BaTi_{1-x}Cr_xAs_2O$ are shown in Figures 4(a) and (b), respectively. Here we take x = 0.25 to facilitate the calculations, although it is higher than the highest doping level of our experiments. Apparently, a significant changes in total DOS can be seen by compared two figures, especially in the region around the Fermi level, where the band width of Cr doped BaTi$_{1.75}$Cr$_{0.25}$As$_2$O is reduced gradually as reference of that of the parent compound BaTi$_2$As$_2$O, resulting in the enhancement of electron correlations being consistent with our magnetism measurements (data not shown here). In addition, we notice that the doping procedure is indeed an introduction of the electron-type carries as Cr doped into the BaTi$_2$As$_2$O as expected from our intuition. The electron dopant leads to the suppression of the charge density wave order, consistent with our transport measurements.

## 3. Conclusion

In the present work, the Cr element was successfully doped into BaTi$_2$As$_2$O. The structure, resistance, and magnetism properties were investigated systematically. The a- and c-axis lattice parameters and volume decrease gradually. Both the transition temperature and the resistive features of the CDW ordering transition are suppressed gradually with the increase of Cr doping. No superconductivity was observed down to 2 K. The clear reduction of the band width and the introduction of electron-type carriers by the Cr doping are evidenced by the first-principles calculations.

**Supporting Information**

Experimental details, details of the first-principles calculations, EDS results and lattice parameters. This material is available free of charge via the Internet at http://pubs.acs.org.

**Acknowledgements**

We acknowledge the help of EDS experiments from Prof. W. Peng and Dr. H. Jin. This work is supported by the Knowledge Innovation Project of Chinese Academy of Sciences (No. KJCX2-EW-W11), the Natural Science Foundation of China (No. 11204338 and 11404359), and the "Strategic Priority Program (B)" of the Chinese Academy of Sciences (No. XDB04040300 and XDB04030000). W.L. also gratefully acknowledges support from the Shanghai Yang-Fan Program (No. 14YF1407100).

# Supporting Information

# Synthesis, Structural, and Transport Properties of Cr-Doped BaTi$_2$As$_2$O

Qiucheng Ji,[a] Yonghui Ma,[a] Kangkang Hu,[a,b] Bo Gao,[a] Gang Mu,[a,*] Wei Li,[a,c] Tao Hu,[a] Ganghua Zhang,[d] Qingbiao Zhao,[d] Hui Zhang,[d] Fuqiang Huang,[d] and Xiaoming Xie[a]

[a] *State Key Laboratory of Functional Materials for Informatics and Shanghai Center for Superconductivity, Shanghai Institute of Microsystem and Information Technology, Chinese Academy of Sciences, Shanghai 200050, China*

[b] *College of Sciences, Shanghai University, Shanghai 200444, China*

[c] *State Key Laboratory of Surface Physics and Department of Physics, Fudan University, Shanghai 200433, China*

[d] *CAS Key Laboratory of Materials for Energy Conversion, Shanghai Institute of Ceramics, Chinese Academy of Sciences, Shanghai 200050, China*

* Corresponding author. Fax: +86-21-62127493. Email: mugang@mail.sim.ac.cn

**Experimental details**

A two-step solid-state reaction method was employed to synthesize the polycrystalline samples of BaTi$_{2-x}$Cr$_x$As$_2$O with the nominal contents 0≤x$_{nominal}$≤0.50. At the first step, CrAs was prepared by reacting Cr and As at 700℃ for 20 h. BaO (99.5%), Ti (99.99%), As (99.995%), and CrAs were then mixed together according to the stoichiometric ratio. The mixture was ground carefully and then pressed into a pellet. The pellet was wrapped by tantalum foil and then sealed in a quartz tube, which was heated at 950℃ for 2 days. Then the samples were reground and sintered again at 950℃ for 3 days. All the weighing and mixing procedures were performed in a glove box with the protective argon atmosphere. The obtained polycrystalline samples demonstrate a dark black color and are sensitive to air.

The actual Cr concentrations were checked and determined by the energy dispersive X-ray spectroscopy (EDS) measurements on a Bruker Quantax200 system. The purity, structure, and lattice parameters were checked by x-ray diffraction (XRD) using a HAO YUAN DX2700 X-ray diffractometer (Cu-$K_{a1}$=1.54051 Å). The resistivity measurements by standard four-probe technique were conducted on a physical property measurement system (PPMS, Quantum Design) down to 2 K. The magnetic properties were measured using the vibrating sample magnetometer (VSM) option based on the PPMS. All the measurements were performed in vacuum or argon atmosphere to protect the samples.

**Details of the first-principles calculations**

The first-principles calculations were performed by using the pseudopotential-based code VASP within the Perdew-Burke-Ernzerhof generalized gradient approximation.[1,2] Throughout the theoretical calculations, a 500 eV cutoff in the plane wave expansion and a 9 × 9 × 6 Monkhorst-Pack grid were chosen to ensure the calculation with an accuracy of $10^{-5}$ eV. The lattice constants were taken from the experimental values shown in Figure S2.

**Elemental analysis**

The actual doping levels were determined by averaging the values obtained on several different spots for each sample. The results are shown in Figure S1. The standard deviations from the averaging process are taken as the error bars. It can be seen that the actual doping levels ($x_{EDS}$) are markedly lower than the nominal ones ($x_{nominal}$), especially in the high doping region. We argue that this is a consequence of solubility limit of Cr in the present system.

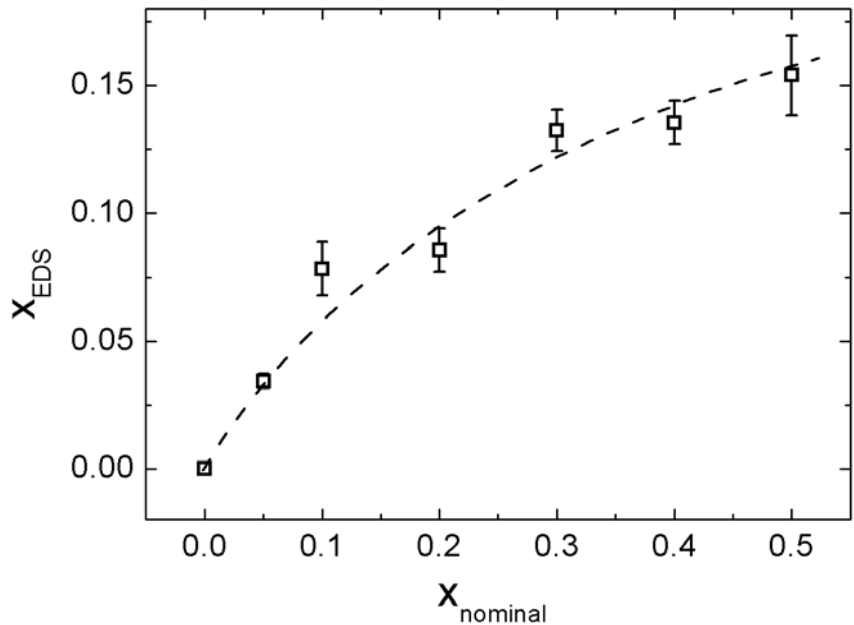

**Figure S1.** The actual doping levels ($x_{EDS}$) determined from the EDS measurements as a function of the nominal values ($x_{nominal}$). The dashed line is the guide for eyes.

**Lattice parameters**

To investigate the influences on the crystal lattice by the Cr substitution quantitatively, we obtained the lattice parameters by fitting the XRD data using the software Fullprof. As shown in Figure S2, the a- and c- axis lattice parameters along with the volume of one unit cell are plotted versus the doping content. The dashed lines are the guides for eyes. The lattice parameters of the undoped sample are quite consistent with the previous work.[3] The data show that the lattice parameters (a- and c- axis) and the volume of one unit cell decrease with the increase of doping. The clear and systematical evolution of such parameters with doping suggests the successful substitution of Cr in the site of Ti.

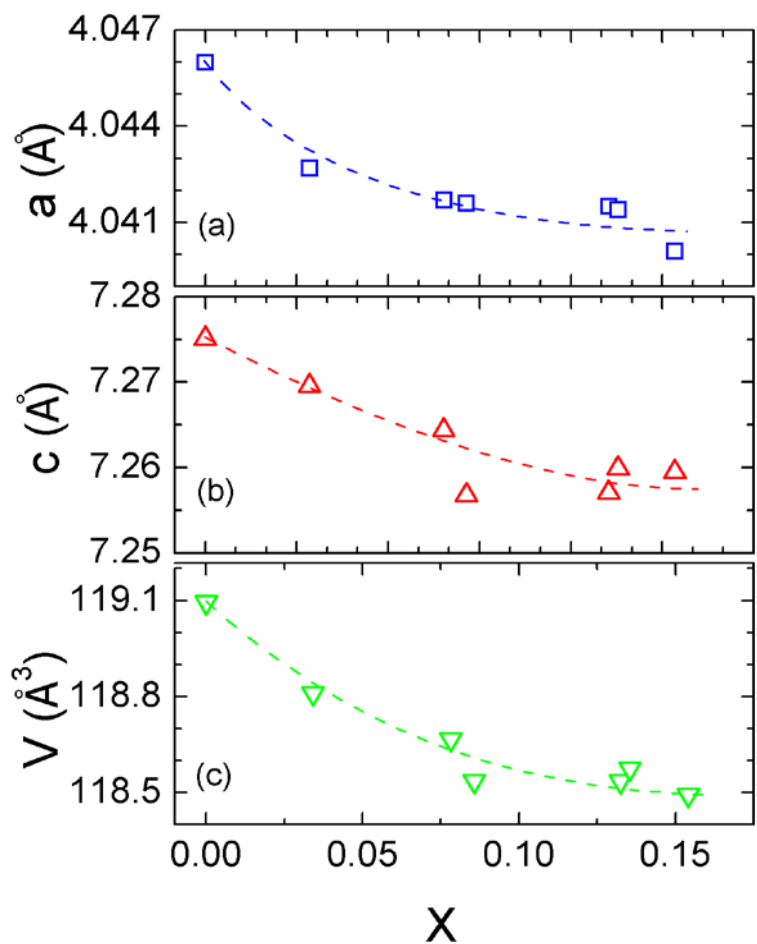

**Figure S2.** Doping dependence of lattice parameters along a-axis (a), c-axis (b), and the volume of one unit cell (c). The dashed lines are the guides for eyes.